\documentclass[12pt,preprint]{aastex}

\usepackage{graphicx}
\usepackage{mathrsfs}
\usepackage{amsmath}
\usepackage{epsfig}
\usepackage{float}        
\usepackage{latexsym}

\def\beq{\begin{equation}}
\def\eeq{\end{equation}}
\def\beqar{\begin{eqnarray}}
\def\eeqar{\end{eqnarray}}

\def\la{\mathrel{\mathpalette\fun <}}

\def\fun#1#2{\lower3.6pt\vbox{\baselineskip0pt\lineskip.9pt
  \ialign{$\mathsurround=0pt#1\hfil##\hfil$\crcr#2\crcr\sim\crcr}}}

\def\qir{q_{\rm IR}}

\begin{document}

\title{Possible Breaking of the FIR-Radio Correlation in Tidally 
Interacting Galaxies}

\author{D. Donevski and T. Prodanovi\'c}
\affil{Department of Physics, University of Novi Sad, \\
Trg Dositeja Obradovi\'ca 4, 21000 Novi Sad, Serbia}
\email{prodanvc@df.uns.ac.rs}

\begin{abstract}
Far-infrared (FIR)--radio correlation is a well-established empirical connection between continuum radio and dust emission of 
star-forming galaxies, often used as a tool in determining star-formation rates. Here we expand the point made by \cite{murphy13}
that in the case of some interacting 
star-forming galaxies there is a non-thermal emission from the gas bridge in between them, which might cause a dispersion in this correlation. 
Galactic interactions and mergers have been known to give rise to tidal shocks and 
disrupt morphologies especially in the smaller of the interacting components. Here we point out that these shocks can also heat the gas and dust 
and will inevitably accelerate particles and result in a tidal cosmic-ray population in addition to standard galactic cosmic rays in the galaxy
itself. This would result in a non-thermal emission not only from the gas bridges of interacting systems, but from interacting galaxies 
as a whole in general. Thus both tidal heating and 
additional non-thermal radiation will obviously affect the FIR-radio correlation of these systems, the only question is how much. 
In this scenario the FIR-radio correlation  is not stable in interacting galaxies, but rather evolves as the interaction/merger progresses.
To test this hypothesis and probe the possible impact of tidal cosmic ray population we 
have analyzed a sample of 43 infrared bright star-forming interacting galaxies at different merger stages. We have found that 
their FIR-radio correlation parameter and radio emission spectral index vary noticeably over different merger stages and behave as it would 
be expected from our tidal-shock scenario. Important implications of departure of interacting galaxies from the FIR-radio correlation are discussed.

\end{abstract}


\section{INTRODUCTION}
\label{sec:intro}

It has long been known that there is a tight correlation between far-infrared (FIR) and radio luminosities in star-forming galaxies 
\citep{vanderkruit71,helou85,condon92,yun01}. This correlation has been shown to hold over almost five orders of 
magnitude for galaxies, not just at local redshifts \citep{yun01}, but also for redshifts from 0 to 2 including various types of galaxies with different 
morphological structures and luminosities
\cite[e.g.][]{ibar08, ivison10,sargent10}.
The underlying physical reason behind the FIR-radio correlation is not fully understood
\cite [see e.g.][]{voelk89,hb93,lacki10,sb13}. It is thought to be due to the 
ongoing star-formation -- dust absorbs light from the massive young stars and emits it in the FIR band, while  
galactic cosmic-ray (GCR) electrons accelerated in supernova remnants emit in radio band.
Consequently, the FIR-radio correlation has been proven to be a powerful tool for determining star-formation rates \citep{condon92}.

Despite numerous studies that have claimed that the FIR-radio correlation is relatively stable \citep{sargent10, bourne11}, 
there are also 
several contemporary observations both at low and high-redshifts that question this.
For example, some observations have shown that the tight 
linear FIR-radio correlation varies in the case of galaxies in rich clusters \citep{ematal09, ry15}, but also for distant starburst 
galaxies \citep{sajina08} 
and distant sub-mm galaxies (SMGs), which were found to be radio brighter with respect to  the local FIR-radio correlation
\citep{kovacs06,magnelli10,smolcic14}.
Though it is generally considered 
\citep{sargent10} that this correlation does not evolve with redshift, the value measured in the few samples
where radio-loud active galactic nuclei have been excluded was found to be more than $10\%$ lower \citep{magnelli10, sajina08}. 
However, what is obvious is the scatter of data around this correlation. This scatter is thought to originate in part from 
1) young active galactic nuclei in which the radio activity 
has begun only recently \citep{drake03}, 2) from much stronger magnetic fields in starbursts  than is suggested by the minimum
energy estimate \citep{thompson06}, or 3) from ongoing interactions between galaxies \citep{murphy13}.

The departure from the typical FIR-radio correlation and excess of radio emission  
was also found in the case of so-called "taffy" systems \citep[interacting galaxies with strong synchrotron-emitting gas 
bridge between them;][]{condon02}. As suggested by \cite{murphy13} this excess in non-thermal emission is probably due to 
particle acceleration in large-scale 
shocks in bridges between interacting galaxies \citep{lisenfeld10}. 
In this work we expand on this and point out that such departure from the typical FIR-radio correlation
may not only be the case for bridges and "taffy" systems, but from cosmic rays accelerated in tidal shocks in the galaxies themselves.

Shock waves that arise during galactic interactions (mergers and close fly-bys) have been known to impact the star-formation 
history of interacting galaxies by triggering star-formation and even leading to a starburst phase \cite[see e.g.][]{sm96}. 
It was pointed out that tidal shocks that accompany these interactions can give rise to a population of 
tidal cosmic rays (TCRs). This can have a potentially significant impact on nucleosynthesis of light elements such as lithium \citep{pub13}. 
Moreover, it is clear that the presence of such cosmic-ray population will 
also result in an enhanced radio emission \citep{pub13}. As a result, the FIR-radio correlation will be impacted.  
This may possible cause the dispersion seen in the relationship and have important implications for star-formation measurements.

In this work we explore the effects that the presence of tidal shocks might have on the FIR-radio correlation in 
interacting star-forming galaxies and test this hypothesis on a small sample of interacting systems. 

\section{FIR-RADIO CORRELATION IN INTERACTING GALAXIES}
\label{sec:form}

The FIR-radio correlation is described using the $ q_{\rm IR}$ parameter \citep{helou85} as
\begin {equation}
\label{eq:qir}
q_{\rm IR}	=\log\left(\frac{F_{\rm FIR}}{3.75\times10^{12}{\rm Wm^{-2}}}\right)-\log\left(\frac{S_{1.4}}{\rm Wm^{-2}Hz^{-1}}\right)
\end {equation}
where $S_{1.4}$ is the continuum radio emission flux at 1.4 GHz per frequency such that 
$S_{1.4} \propto \nu^{-\alpha}$ and $\alpha$ is the radio spectral index, positive in vast majority of sources. 
$F_{\rm FIR}$ is the rest-frame FIR dust emission flux.
\cite{yun01} analyzed the sample of 1800 {\it IRAS} 
galaxies and measured this value to be  $q_{\rm IR}=2.34\pm0.01 $  with a dispersion of 0.25 dex. Both normal star-forming spirals 
and merger-induced luminous infrared galaxies were included in their sample. 

In this work we revisit the idea of galactic interactions being (in part) the source of the large dispersion of the relationship.
As galactic interactions can give rise to tidal shocks in the interstellar medium of 
interacting galaxies, they can
consequently produce  a cosmic-ray population, in addition to normal GCRs, resulting in excess radio emission. 
Such an enhancement in radio flux would cause the dispersion of the FIR-radio correlation,
which would especially be significant at higher redshifts with increased rate of galactic interactions compared to low-redshift Universe.

Unlike GCRs, cosmic rays accelerated in tidal shocks due to close galactic fly-bys 
would result in 
an excess of non-thermal emission in gamma-rays and radio band that is not immediately 
accompanied by the corresponding increase in star-formation rate \citep{pub13}. 
Therefore the most promising way for identifying the presence of 
TCRs is to 
observe galaxies in early (mid) merging stages \citep[merging stage 3 according to][classification 
scheme]{haan11}.
One would also anticipate additional heating of the gas and especially dust in tidal shocks
\citep[as was observed in interacting systems; see e.g.][]{lanz13,mc12}, compared 
to what would normally be expected due to the ongoing star-formation, again, provided that the system is observed in an 
sufficiently early merging stage.
With these additional sources of non-thermal and thermal radiation it is clear that the FIR-radio 
correlation would be affected such that $q_{\rm IR}\ne q_{\rm IR}^{\rm T}$ -- a typical FIR-radio parameter
will not be equal to that same parameter $ q_{\rm IR}^{\rm T}$ for a system where there is a presence of additional cosmic-ray population 
(such are for example TCRs). 
More specifically, if there are tidal shocks and a TCR population present, then the expected parameter would be 
\beq
q_{\rm IR}^{\rm T}=\log \frac{F_{\rm FIR}+F_{\rm FIR}^{\rm TS}}{S_{1.4}+S_{1.4}^{\rm TS}}.
\eeq
The $F_{\rm FIR}^{\rm TS}$ is the additional FIR flux of dust coming from tidal shock heating, while $S_{1.4}^{\rm TS}$ is 
the additional radio flux from TCR electrons. Observing that $q_{\rm IR}>1$ and assuming that the effects of tidal shocking
are a small perturbation already existing effects (i.e. that $F_{\rm FIR}^{\rm TS}/S_{1.4}^{\rm TS}<F_{\rm FIR}/S_{1.4}$) we will
have that in general 
$q_{\rm IR}> q_{\rm IR}^{\rm T}$. However, as the interaction between galaxies progresses, how would we
expect this parameter to change?

\subsection{Early Interaction: Enhanced Heating}

At the very early stages of interaction,  we expect tidal shocks to form in the ISM and start heating 
the dust and gas. Dust can be heated in collisions with gas or by shock UV radiation, 
causing it to radiate thermally in infrared.  However, this will also result in destruction of dust
 due to sputtering processes. The  time-scale of destruction of dust in shocks goes from few thousand
years to tens of millions of years \citep{tielens98,lau2015,villar-martin,dwekarendt}, depending mainly on the grain size, strength of the shock 
and density of the ambient medium. 
On the other hand typical cosmic-ray acceleration timescale in supernova remnants is
of the order of lifetime of the remnant, being a few $\sim 10^5$ yr \citep[e.g.][]{strong2007}. 
For large-scale shocks considered here, cosmic-rays will
be accelerated as long as the tidal shock propagates, which can be of the order of Gyrs.
Thus, during the first few thousand to million years of galactic interaction, we can consider that there is enhanced thermal emission of heated dust and gas over what is 
typically expected of non-interacting systems, without enhanced non-thermal emission of freshly accelerated cosmic-rays. 
As a result, we can expect a non-thermal radio emission dominated by the already present GCR population, and thermal emission with an added contribution from
tidal shock heating, leading to $q_{\rm IR}< q_{\rm IR}^{\rm T,h}$.
The enhanced overall spectral index will also change, becoming shallower (harder)  $\alpha < 0.8$, 
compared to a typical observed value $\alpha \approx 0.8$ \citep{condon92}. A shallower radio spectral index indicates a more
dominate thermal component \citep[see e.g.][]{lisenfeld00}, and vice versa. 
We expect this change in spectral index to accompany the change in the FIR-radio parameter on a same timescales.  

It is generally accepted that galaxies like the Milky Way are not to be considered as closed-box systems (unlike starburst galaxies), but should be expected to
"leak" cosmic-rays on timescales of $\tau_{\rm esc} \sim 2 \times 10^7$ yr \citep{garcia-munoz77}. So for a galaxy like the Milky Way, with supernova rate of
$\dot{R}_{\rm SN}\sim 1/50$ yr \citep{tammann94}, $4 \times 10^5$ supernova events will occur before 
escape losses become important, allowing an estimate of the maximal GCR flux. In order for tidal shocks to result in a cosmic-ray flux 
greater than the already present GCR flux, their 
input needs to be equivalent to about $10^5$ supernova events. Consequently, the volume of gas shocked by tidal shocks needs to be equal to volume shocked by that many
supernova events, assuming the same acceleration efficiency. 
Taking that particles are efficiently accelerated in supernova remnants up to the radius of $R_{\rm SN} \sim 10$ pc 
\citep{berezhko}, we find that tidal shocks would need to shock all the gas up to the radius of about 1 kpc in a galactic disk of thickness $d=300 $pc. 
This would take about 10 million years, assuming a conservative radial tidal shocks with velocity $v_{\rm shock}\sim 100 \rm km/s$ 
\citep[tidal shocks are weaker than collision shocks which are of the order of the velocity of merging galaxies $\sim $ few hundred km/s; see e.g.][]{kashiyama2014}. 
Therefore, we estimate that this early stage where non-thermal emission due to TCRs can be neglected, would last during 
the first few million of years, or more precisely
\beqar
\nonumber
\tau_{\rm early} & \la & 13 {\rm Myr} \left( \frac{v_{\rm shock}}{100 {\rm km/s}}\right)^{-1} \left( \frac{R_{\rm SN}}{10 {\rm pc}}\right)^{3/2}  
\left( \frac{d}{300 {\rm pc}}\right)^{-1/2} \\
&& \times \left( \frac{\tau_{\rm esc}}{2 \times 10^7 {\rm yr}}\right)^{1/2} \left( \frac{\dot{R}_{\rm SN}}{1/50 {\rm yr}} \right)
\eeqar 

In the case of starburst galaxies with $\sim 10 \times$ higher supernova rates, this timescale would be about a factor of 3 higher.
Note however that supernova (i.e. star-formation) rate determined from radio observations would not be suitable within the framework of our
hypothesis because if tidal cosmic-ray population is present it might lead to enhanced radio emission resulting in a overestimate of star-formation rates.
A better way, free from possible "contamination", would be to use star-formation rates determined from H$\alpha$ observations to test how galactic interactions might impact our understanding of star-formations rates. This will be the topic of a follow-up study.

\subsection{Mid Interaction: Enhanced Particle Acceleration}

At few tens of million of years since the beginning of interaction, acceleration of particles in tidal shocks
will start to be significant. In an extreme scenario, non-thermal emission of a population of tidal cosmic-rays
will at some point become dominant over the galactic cosmic-ray population. This would be late enough in the interaction so that
the star-formation has been triggered by tidal shocks, resulting in
additional heating of the gas and dust due to new stars. 
The approximate start of this stage would be at the timescale of stellar contraction and formation, that is, this 
stage would
begin at least tens of millions of years after the formation of tidal shocks. 

Tidal shock heating then becomes less important as the source of thermal emission.
On the other hand, TCRs are now a dominant source of non-thermal emission given that formation of new stars has not yet been accompanied with 
the increase in GCRs flux.
In that case we expect a decrease in the FIR-radio parameter below its typical value $q_{\rm IR}^{\rm T,cr} < q_{\rm IR}$,
such that $q_{\rm IR}^{\rm T,h}>q_{\rm IR}> q_{\rm IR}^{\rm T} > q_{\rm IR}^{\rm T,cr}$.
The decrease in the FIR-radio parameter would be accompanied with a steeper (softer) radio spectral index $\alpha > 0.8$, compared 
to a typical observed value in radio, reflecting a larger presence of a non-thermal component. Again, this
evolution in spectral index would be expected to be on the same timescale as the evolution of the FIR-radio
parameter.

A quick look into energetics also leads to the conclusion that tidal cosmic-ray population may become
dominant at some point. Namely, one supernova event injects approximately $\dot{E}_{\rm SN}=10^{51}/50=2 \times 10^{49}$ erg/yr into the
interstellar medium, where $\sim 10\%$ of that energy is generally considered to go into particle
acceleration. We note that this energy-injection rate is valid for normal, star-forming galaxy like the Milky
Way. In the case of starburst galaxies with much higher supernova rates, energy-injection rate would clearly also be higher.
This means that the following estimate is valid for the early epoch of galactic interaction, where supernovae from subsequently triggered 
starburst phase have not yet injected their energy. That will be the epoch where tidal cosmic-ray flux can be a potentially
significant addition to the already present galactic cosmic-ray flux.  
Following \cite{pub13} we estimate the kinetic energy of the encounter  between
 Milky Way type galaxies at a distance of $50$ kpc is $E_{\rm maj}=4 \times 10^{60}$ erg. For a minor merger where smaller 
component is 1000 times less massive energy of interaction is $E_{\rm min}=4 \times 10^{57}$ erg. 
This energy estimate is of the order of tidal energy that gets injected in the system \citep[see e.g.][]{spitzer58}.
If $10\%$ \citep[see e.g.][]{kashiyama2014} of this energy goes into particle acceleration and if interaction timescale is of the order of $10^9$ yr, 
we find that the rate at which tidal shocks can inject energy into particles is $\dot{E}_{\rm T,maj}=4 \times 10^{50}$ erg/yr for major merger and 
$\dot{E}_{\rm T,min}=4 \times 10^{47}$ erg/yr for minor merger. We see that in the case of major mergers, tidal shocks have the potential
to dominate the non-thermal emission. In the case of minor mergers TCRs can be at the $10\%$ level of the GCRs, which is a smaller effect. Note however that this was estimated assuming that the smaller component has large star-formation and  supernova rates equal to the Milky Way. In reality, for a smaller system, we would expect this rate to be orders of magnitude lower for a galaxy 1000 times less massive than Milky Way. In that case energy injection rate of tidal shocks compared to supernovae would be much higher.  On the other hand, in later phases, when star-formation is triggered by the merger, star-formation rate of such small systems can indeed be as high as for the Milky Way.

Another way to look into energetics is to compare the energy of tidal interaction with the non-thermal luminosity of the source. 
Let us, for example, consider the case of stage 3 system NGC5256, with component masses of $6.3 \times 10^{10} M_\odot$ and 
$5.3 \times 10^{10} M_\odot$  and separation of about 7.5 kpc \citep{mazz12}. If we then assume that they tidally interact over the timescale of 1 Gyr, 
we find that tidal energy injection rate is of the order of $\dot{E}_{\rm T}= 3 \times 10^{42}$ erg/s. If we compare this to source 
luminosity at 4.8 GHz $\nu L_\nu = 4.1 \times 10^{39}$ erg/s \citep{mazz12} we see that for
electron-to-proton ratio of 1/100 at energy 1 GeV, $ \sim 10\%$ of tidal interaction energy needs to be converted into 1 GeV 
cosmic-rays to account for luminosity at 4.8 GHz band. Acceleration efficiency for even this conservative estimate is 
consistent with typical acceleration efficiencies in supernova remnants \citep{lisenfeld10}.

In order to check if TCRs accelerated in large-scale tidal shocks that traverse through entire interacting galaxies, can produce observed non-thermal
radio emission and spectra, we analyze the case of IC1623, which is one of the systems in our sample that is a stage 3 merger with lower than typical FIR-radio parameter $q_{\rm IR}=2.08$, and steeper than typical spectral index $\alpha=0.91$. Following Bell's theory of particle acceleration in shock fronts \citep{bell78} we calculate radio emissivity of shocked gas assuming that the entire gas of IC1623 was shocked by tidal shocks of speed 250-300 km/s, that number density
of ISM is 1 $\rm cm^{-3}$ and that the average galactic magnetic field is of strength 10-25 $\mu$G \citep{Drzazga11,beck09}. Based on that, and adopting its measured spectral index we calculate the expected luminosity at 1.4 GHz $L_{1.4}=1.3-4 \times 10^{23}$ W/Hz. This is in very good agreement with luminosity calculated from observed flux $L_{1.4}=2.1-4 \times 10^{23}$ W/Hz \citep{murphy13}.

\subsection{Late Interaction: Enhanced Star-Formation}

Tens to hundreds of million of years from the beginning of interaction, star-formation rates would be increasing, accompanied by the
increase in supernova rate but with slight offset at order of lifetime of massive stars. 
This is consistent with results from numerical models \citep[e.g.][]{dimatteo2008} which give that first enhancements in star-formation rates would be
few hundred million years in the interaction after the first passage, followed by the star-burst epoch at time $\sim 1$ Gyr from the beginning of 
interaction and lasting some 500 Myrs. During that time the FIR-radio parameter would start to grow as FIR emission would be 
increasing due to new stars being born.
Finally, at the end of interaction one would expect a similar epoch of enhanced supernova rate. At that point, the FIR-radio parameter would be 
expected to go back to its typical value comparable to that of isolated systems. 
Finally, we note that at higher redshifts, timescales and durations of each phase would be different due to lower concentration of dust
and lower metallicity \citep[e.g.][]{pettini1997}. 

\begin{table}[htpb]
\begin{center}
\epsscale{1.0}
\caption{Merger stage classification according to \citet{haan11} with corresponding symbols used on plots.}
\label{table:stages}
\begin{tabular}{ccl}
\hline
Merger & Symbol & Description  \\
stage & & \\
\hline
0 & None & Non-merger \\
1 & $\bigtriangleup$  & Pre-merger: separate \\
 & & galaxies, non tidal tails \\
2 & $\Diamond$ & Ongoing merger, early: \\
 & & progenitor galaxies distinguishable  \\
3 & $\bullet$ & Ongoing merger: \\
 & & progenitors sharing an envelope \\
4 & $\star$ & Ongoing merger, late:  one galaxy \\
 & & with double nuclei and a tidal tail\\
5 & $\Box$ & Post-merger: one galaxy with single \\
 & & (disturbed) nucleus and prominent tails \\
6 & None & Post-merger, late: one galaxy \\
 & &  with single nucleus and weak tail \\
\hline
\end{tabular}
\end{center}
\end{table}

\section{Analysis of The Sample of Interacting Galaxies}

As described, what would eventually be expected from galactic interactions is the dependance of the FIR-radio parameter $q_{\rm IR}$ on the 
merger stage. Early in the interaction this parameter would have larger than the average value for the non-interacting star-forming 
galaxies, after which it
would be expected to decrease to lower than the average value, following eventual relaxation back to the original value. 
To test this we have analyzed the 
sample of 43 IR bright galaxies found in  
data sets from \cite{dopita02} and \cite{murphy13}. Data presented in \cite{murphy13} were drawn from {\it IRAS} 
revised Bright Galaxy Sample \citep{sanders03}. Galaxies in this sample were chosen to  have $60\mu \rm m$ flux densities 
larger then 5.24 Jy 
and FIR luminosities $\geq10^{11.25}L_{\odot}$. Taking into account that original \cite{dopita02} data set consists of 
two studies \citep{kewley01,corbett02} where different lower limit for  FIR 
luminosities was chosen, our sample used in this work satisfies the overall criteria that all chosen galaxies are IR bright with 
$60\mu$m 
flux densities larger than 2.5 Jy and have IR luminosities $\geq10^{10.5}L_{\odot}$ . We have excluded all sources classified as 
active galactic nuclei in the original data sets. All galaxies presented here have well sampled radio spectra between 1.4 and 8.4 GHz, but for most of them  data is also available outside of that range of frequencies.
In total 15 galaxies from \cite{murphy13} are used. Values of $\qir$ were taken directly from \cite{murphy13}. For \cite{dopita02} sample we have calculated $\qir$ following equation (1), estimating the far-infrared flux, ${F_{\rm FIR}}$, as defined in \cite{sm96}, based on {\it IRAS} flux densities
$f_{\nu}$ at $60\mu \rm m$ and $100\mu \rm m$:
\beq
\left(\frac{F_{\rm FIR}}{\rm Wm^{-2}}\right)=1.26\times10^{-14}\left[\frac{2.58f_{60\mu \rm m}+f_{100\mu \rm m}}{\rm Jy}\right].
\eeq

For galaxies from \cite{murphy13} radio spectral indices were recalculated for archival {\it VLA} data at 1.4 and 4.8 GHz from standard 
$\alpha=\log\left(S_{1.4}/S_{4.8}\right)/\log\left(4.8/1.4\right)$. Galaxies studied in \cite{dopita02} are COLA galaxies \citep{corbett02}, and their indices were determined at same frequencies as in the original \cite{corbett02} paper. Merger stages have been defined following 
\cite{haan11}. Classification status for each galaxy was taken from \cite{murphy13} and \cite{dopita02}. We have excluded galaxies unclassified by their merging stage in \cite{murphy13}, while for one of them (NGC 6286S) we  have additionally confirmed (inspecting the optical DSS images on the NASA/IPAC Extragalactic Database - NED) its merging stage 3 according to apparent weak tidal features and an envelope shared with its interacting companion. 

\begin{figure}
\begin{center}
\epsscale{1.0}
	\includegraphics[width=74mm]{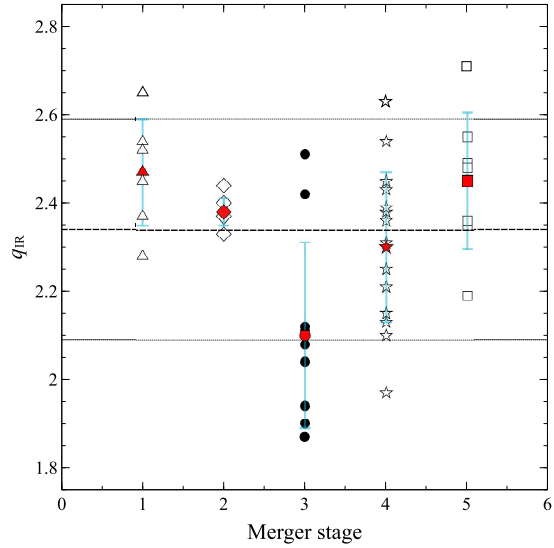}
	\caption{Plotted is the FIR-radio correlation parameter $q_{\rm IR}$ for the sample of 43 interacting star-forming galaxy pairs at 
different merger stages. Merger stages \citep{haan11} were labeled using the same symbols as noted in Table 1. 
The dashed line 
represents the typical value of the $q_{\rm IR}$ \citep{yun01}, while dotted lines represent its mean deviation. Mean values and their standard deviations were calculated for each merger stage subsample separately and are presented with red symbols and blue error bars.}
	\label{fig:qvsstage}
\end{center}
\end{figure}

We have also calculated the mean values and standard deviations of $q_{\rm IR}$ parameter and spectral index $\alpha$ separately for all interaction stages of 
galaxies in our sample. This is presented in Table 2. 
On Figure \ref{fig:qvsstage} we plot FIR-radio parameter $q_{\rm IR}$ as a function of the merger stage for this sample of IR 
bright interacting galaxies. The original idea that the FIR-radio parameter $q_{\rm IR}$ is sensitive to merger stage was presented 
in \cite{murphy13}. Figure 5 from \cite{murphy13} revealed significant scatter of the FIR-radio parameter for sources at different merger stages, 
and here, we have improved the statistics with additional data set of IR bright sources to 
see is the same behavior of $q_{\rm IR}$ related to merger events will hold. Though statistics is still limited, we see that 
only in the case of stage 4 and 5 mergers is the mean value of the FIR-radio parameter consistent with its global value of 
$q_{\rm IR}=2.34$ \citep{yun01}. The average $q_{\rm IR}$ for the other merger stages deviate from the global value. 

For example, for the most significant case of merger stage 3 (where we expect that the impact of tidal cosmic-ray 
population is most prominent), we perform a statistical t-test to check how significant the offset of this data subset is 
is from the sample mean value of $q_{\rm IR}=2.34 \pm 0.21$. We find a t-value of 
$t=3.8$ corresponding to a p-value of $p=0.004$, indicating that the probability that this offset is random is less than $1\%$. 
This offset is even more significant if we compare it to the population mean with $q_{\rm IR}=2.34 \pm 0.01$.
 What it appears is that systems 
 in close fly-bys, in the pre-merger stages, have higher than average values of $q_{\rm IR}$, which then decreases toward later 
 merger stages and reaches the minimum value at merger 3 systems, after which it increases again.
We note that \cite{murphy13} has shown similar trend plotting the difference between the observed and nominal logarithmic IR 
and radio flux densities, but without any further statistical analysis.
 We should point out that two of merger stage 3 galaxies show "no excess" to standard $q_{\rm IR}$ value.
However, inspecting NED images of those sources 
 reveals that source IRAS 06295 shows no obvious tidal features, which is a typical property of stage 3 mergers. The other galaxy IRAS 12286-2600 has a smaller companion with double nuclei, which is again not typical of a simple stage 3 system. 

\begin{table}
\begin{center}
\epsscale{1.0}
\caption{FIR-radio parameter $q_{\rm IR}$ and spectral index $\alpha$ values calculated for each merger-stage subsample separately.}
\label{table:values}
\begin{tabular}{cccc}
\hline
Merger & Symbol & $q_{\rm IR}$ & $\alpha$  \\
\hline
1 & $\bigtriangleup$  & 2.47 $\pm$ 0.12 &  0.69 $\pm$ 0.171 \\
2 & $\Diamond$ & 2.38 $\pm$ 0.03 & 0.56 $\pm$ 0.185\\
3 & $\bullet$ & 2.09 $\pm$ 0.21 & 0.92 $\pm$ 0.19 \\
4 & $\star$ & 2.31 $\pm$ 0.17 & 0.79 $\pm$ 0.22 \\
5 & $\Box$ & 2.44 $\pm$ 0.15 & 0.51 $\pm$ 0.16 \\
\hline
\end{tabular}
\end{center}
\end{table}

\begin{figure}
\begin{center}
\epsscale{1.0}
	\includegraphics[width=74mm]{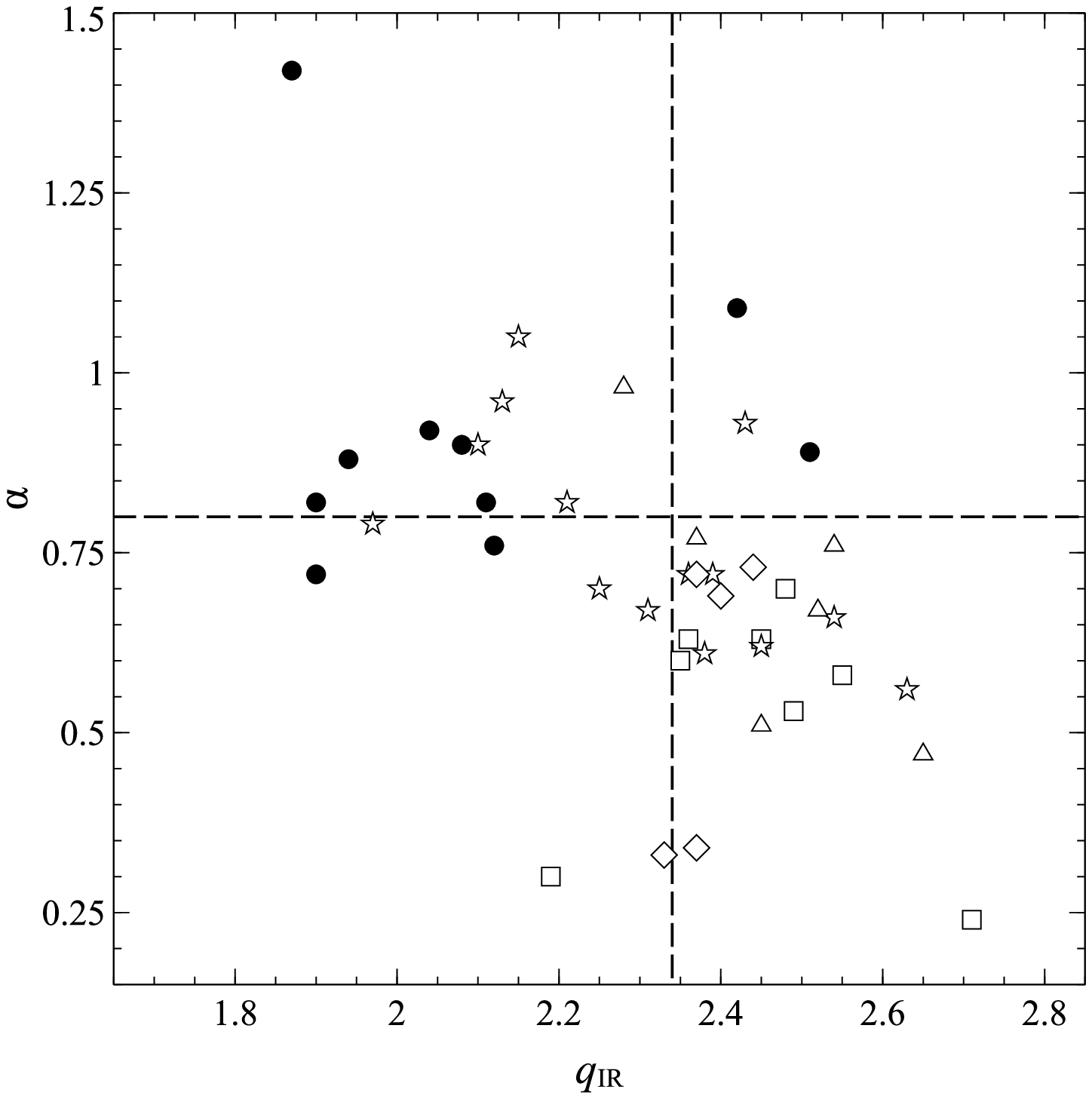}
	\caption{Plotted is the radio spectral index $\alpha$ ($S_{1.4} \propto \nu^{-\alpha}$) measured between 1.4 GHz and 4.8 GHz,
 as a function of the $q_{\rm IR}$ parameter for the same sample of interacting star-forming galaxies at different merger stages \citep{haan11} labeled using the 
same symbols as noted in Table 1. Dashed lines represent typical values of radio spectral index $\alpha = 0.8$ \citep{condon92} and 
$q_{\rm IR}$ \citep{yun01} parameter for star-forming galaxies. For our data set we have calculate the scatter around the mean spectral index to be 
$\alpha = 0.8 \pm 0.05$ which is consistent with results from \cite{murphy13}, while the scatter of the mean FIR-radio parameter is $q_{\rm IR}=2.34 \pm 0.21$.}
	\label{fig:alphavsq}
\end{center}
\end{figure}

As was discussed in Section 2, the evolution of the FIR-radio parameter $q_{\rm IR}$ with respect to the merger stage would also be accompanied with the 
corresponding evolution in the radio spectral index. Ongoing mid-stage mergers would be expected to have enhanced non-thermal emission with $\alpha_{\rm T} > 
\alpha=0.8$ and $q_{\rm T}<q=2.3$. This trend is actually seen on Figure \ref{fig:alphavsq} where we have plotted radio 
spectral index $\alpha$ as a function of the FIR-radio parameter $q_{\rm IR}$. The upper left quadrant of the plot where $q_{\rm IR}$ 
is lower than typical $\alpha = 0.8 \pm 0.05$ (dashed line) and radio spectral index is higher than typical (dashed line), is dominated by merger stage 3  and 4 systems. 
As opposed to that, the bottom right quadrant with $q_{\rm IR}$ 
higher than typical  and radio spectral index lower than typical value, is dominated by the
late merger stages 4 and 5, as would be expected. In the case of merger stage 3 the t-test gives a p-value of 
$p=0.03$ that this offset of
stage 3 spectral index from typical value is random.

\section{DISCUSSION AND CONCLUSION}
\label{sec:disc}

In this paper we explore and draw attention to important effects that close-galactic interaction and mergers can have 
on the stability of the widely-used FIR-radio correlation. 
Close galactic interactions produce tidal shocks in interacting galaxies which
leads to gas and dust heating. Moreover, particle acceleration in tidal shocks gives rise to a tidal cosmic-ray population. These two effects would 
impact the infrared and radio emission of the interacting galaxies and cause variations in the FIR-radio parameter $\qir$ and 
radio spectral index over different merger stages. To test this, we have analyzed the sample of 43 IR bright interacting galaxies in 
different merger stages, looking at how their FIR-radio parameter and radio spectral index change over different merger 
stages. 
What we have tentatively found is that $\qir$ first decreases during early merger stages, and then later increases. This is consistent with what 
would be expected if the heating of the gas and dust in tidal shocks is the dominant effect at early merger stages, followed by the phase 
where tidal cosmic-ray emission dominates over the existing GCR emission, and eventually ending with the enhanced star-formation taking this correlation parameter
back to its typical value. As a result, if interacting galaxies are included in the FIR-radio correlation, they could be one possible cause of its
observed dispersion, and probably a dominant cause of its dispersion going to a higher redshifts where merger rates are higher.

Our results are consistent with results of \cite{murphy13} who was the first to point out the possible impact of galactic interactions on the FIR-radio relation. While \cite{murphy13} was mostly focused on describing the radio emission of "Taffy"-like systems, here we wanted to 
also include the cases where there is amplified radio emission like in the galaxy NGC5256 and galaxies alike which cannot be treated as "Taffy"-like due to lack of $\rm H_2$ gas in 
the bridge between the two galaxies. More specifically, the NGC5256 consists of a pair of galaxies, comparable in size and scale \citep{tigran80} with projected 
separation between their nuclei of about 8 kpc. It has a very low $q_{\rm IR}=1.90$ value, and recent multi-wavelength 
studies \citep{mazz12} uncovered several interesting features in these luminous IR-galaxies:
(a) The optical morphology of the NGC 5256 north-east nuclear environment is similar to the radiative shock observed south 
of the nucleus of M51; (b) Bridge of CO$\left(1-0\right)$ emission is spatially decoupled from the radio continuum emission; (c) 
Steeper radio index is observed not only in the region in between galaxies, but also close to the edges of both components 
\citep[see radio maps in][]{var15}; (d) Whereas the bulk of the $\rm HCO^+$ molecules in Taffy systems is located between 
the nuclei, the $\rm HCO^+$ in NGC5256 is still bound to the galaxies; (e) Spectral energy distribution (SED) of NGC 5256 can 
be modeled as a starburst-dominated, however compared to galaxies with the same SED description (for example, NGC 2623 
or NGC 6240), but different merging stage, the NGC5256 shows evidence of higher dust temperatures; (f) a soft X-ray emission 
extending 15 kpc to the north of the system, between the nuclei, was observed \citep{brassington07}, revealing the presence 
of shock-heated gas indicating that corresponding synchrotron radio emission is shock-induced. This case shows that even in non-"Taffy"' systems 
there can be important shocking of gas and dust and particle acceleration within the interacting galaxy.	
Furthermore, \cite{Drzazga11}  
have studied how tidal interactions affect the evolution of galactic magnetic fields and according to their analysis of polarization data 
in two Taffy systems (UGC12914/5 and UGC813/6), they found a well-ordered magnetic field in the bridge in one of them (UGC12914/5), 
corresponding more to a shocked gas well before a adiabatic (Sedov) phase, which was the scenario analyzed in \cite{lisenfeld10} to explain the bridge emission. 
All of this indicates that tidal shocks in interacting galaxies, their evolution and effects can provide a general solution to effects observed in interacting and merging systems.
 
If tidal shocks and cosmic-ray acceleration are the underlying reason for merging and interacting galaxies to deviate from 
the well-established FIR-radio correlation, this effect can be used at high redshifts as a tool in searching for interacting systems and testing our 
understanding of high redshift interaction rates. Moreover, departure from the FIR-radio correlation in the case of interacting 
systems could have important consequences for determination of star-formation rates 
\citep{yuncarilli02,bell03,carilli08,dunne09} leading to its overestimate. A more reliable way to determine
star-formation rates in interacting galaxies would be to look into H$\alpha$ emission that directly probes emission of massive young stars. 
For example, the mean star-formation rate determined from radio observations for our entire sample of 43 galaxies is $41 M_\odot$/yr, while
H$\alpha$ observations give the average of $16 M_\odot$/yr. The effects of this will be explored in the follow-up work.

In order to determine if indeed interacting systems deviate from the well-established FIR-radio correlation and how, a larger 
sample of bright IR galaxies along with radio data has to be analyzed. The data available from surveys such are 
${\it COSMOS}$ and  ${\it CANDELS}$ along with the upcoming data from {\it ALMA} will provide a perfect testing ground for 
this. 
For example, ${\it CANDELS}$ survey is using deep near-IR imaging to reveal morphological classifications of distant 
IR-bright galaxies and directly counts the number of interacting pairs up to ($z\sim2$), while ${\it COSMOS}$ survey uses
interferometric follow-up observations of distant dusty star-forming galaxies to redshifts even greater than $z\sim4$. 
With {\it ALMA} it would be possible to resolve these dusty galaxies into individual pairs for a large sample of radio faint 
submillimeter galaxies (SMGs). Combining different radio maps it would be possible to determine spectral indices of distant SMGs, and trace the FIR-radio correlation for a large sample of SMGs free from the biases. 

Besides the need for a larger sample, it would also be important to obtain multi-wavelength observations of systems that are found to be good candidates to be dominated by tidal cosmic ray population. An example of such systems would be the Whirlpool galaxy, M51 (or NGC 5194) and NGC 5256. Recent PdBI Arcsecond Whirlpool Survey (PAWS) \citep{eva13} revealed the presence of additional cosmic ray emission in spiral arms of the NGC 5194 where low rate of ongoing star formation is present. The smaller companion, NGC 5195, was found to have low  or no ongoing star formation \citep{bigiel08}, however, it has uncharacteristically high dust temperature \citep{mc12}, possibly due to shock heating.

Since the effects of tidal shocks would be most pronounced in the smaller of the interacting components
for which very little data is currently available, it would also be important to observe best candidates
of those smaller galaxies especially in radio and IR domains.  

\subsection*{Acknowledgments}

We are grateful to Francois Mernier and Jovana Petrovic for valuable discussions and comments. We are also thankful to the anonymous Referee for
very constructive comments and points which improved this paper. The work of TP is supported in
part by the Ministry of Science of the Republic of Serbia under project numbers 171002 and 176005.

\label{lastpage}


\begin{thebibliography}{}

\bibitem[Beck(2009)]{beck09} Beck, R.\ 2009, Astrophysics and Space Sciences Transactions, 5, 43 

\bibitem[Bell(2003)]{bell03} Bell, E.~F.\ 2003, ApJ, 586, 794

\bibitem[Bell(1978)]{bell78} Bell, A.~R.\ 1978, MNRAS, 182, 443 

\bibitem[Berezhko \& V\"{o}lk (2004)]{berezhko} Berezhko, E.~G., V\"{o}lk, H.~J.\ 2004, A\&A, 427, 525 

\bibitem[Bigiel et al.(2008)]{bigiel08} Bigiel, F., Leroy, A., Walter, F., et al.\ 2008, AJ, 136, 2846

\bibitem[Bourne et al.(2011)]{bourne11} Bourne, N., Dunne, L., Ivison, R.~J., et al.\ 2011, MNRAS, 410, 1155 

\bibitem[Braine et al.(2003)]{braine03} Braine, J., Davoust, E., Zhu, M., et al.\ 2003, A\&A, 408, L13 

\bibitem[Brassington et al.(2007)]{brassington07} Brassington, N.~J., Ponman, T.~J., Read, A.~M.\ 2007, MNRAS, 377, 1439

\bibitem[Carilli et al.(2008)]{carilli08} Carilli, C.~L., Lee, N., Capak, P., et al.\ 2008, ApJ, 689, 883 
  
\bibitem[Condon(1992)]{condon92} Condon, J.~J.\ 1992, ARA\&A, 30, 575 

\bibitem[Condon et al.(2002)]{condon02} Condon, J.~J., Helou, G., \& Jarrett, T.~H.\ 2002, AJ, 123, 1881

\bibitem[Corbett et al.(2002)]{corbett02} Corbett, E.~A., Norris, R.~P., Heisler, C.~A., et al.\ 2002, ApJ, 564, 650 

\bibitem[Di Matteo et al.(2008)]{dimatteo2008} Di Matteo, P., Bournaud, F., Martig, M., et al.\ 2008, A\&A, 492, 31 

\bibitem[Dopita et al.(2002)]{dopita02} Dopita, M.~A., Pereira, M., Kewley, L.~J., \& Capaccioli, M.\ 2002, ApJS, 143, 47

\bibitem[Drake et al.(2003)]{drake03} Drake, C.~L., McGregor, 
P.~J., Dopita, M.~A., \& van Breugel, W.~J.~M.\ 2003, AJ, 126, 2237 

\bibitem[Drzazga et al.(2011)]{Drzazga11} Drzazga, R.~T., Chy{\.z}y, K.~T., Jurusik, W., \& Wi{\'o}rkiewicz, K.\ 2011, A\&A, 533, A22 

\bibitem[Dunne et al.(2009)]{dunne09} Dunne, L., Ivison, R.~J., Maddox, S., et al.\ 2009, MNRAS, 394, 3 

\bibitem[Dwek \& Arendt(1992)]{dwekarendt} Dwek, E., \& Arendt, R.~G.\ 1992, ARAA, 30, 11 

\bibitem[Garcia-Munoz et al.(1977)]{garcia-munoz77} Garcia-Munoz, M., Mason, G.~M., \& Simpson, J.~A.\ 1977, ApJ, 217, 859 

\bibitem[Haan et al.(2011)]{haan11} Haan, S., Surace, J.~A., Armus, L., et al.\ 2011, AJ, 141, 100 

\bibitem[Helou et al.(1985)]{helou85} Helou, G., Soifer, B.~T., \& Rowan-Robinson, M.\ 1985, ApJL, 298, L7 

\bibitem[Helou \& Bicay(1993)]{hb93} Helou, G., \& Bicay, M.~D.\ 1993, ApJ, 415, 93 

\bibitem[Ibar et al.(2008)]{ibar08} Ibar, E., Cirasuolo, M., Ivison, R., et al.\ 2008, MNRAS, 386, 953 

\bibitem[Ivison et al.(2010)]{ivison10} Ivison, R.~J., Magnelli, B., Ibar, E., et al.\ 2010, A\&A, 518, LL31 

\bibitem[Kashiyama 
\& M{\'e}sz{\'a}ros(2014)]{kashiyama2014} Kashiyama, K., \& M{\'e}sz{\'a}ros, P.\ 2014, ApJL, 790, L14 

\bibitem[Kewley et al.(2001)]{kewley01} Kewley, L.~J., Heisler, C.~A., Dopita, M.~A., \& Lumsden, S.\ 2001, ApJS, 132, 37 

\bibitem[Kov{\'a}cs et al.(2006)]{kovacs06} Kov{\'a}cs, A., Chapman, S.~C., Dowell, C.~D., et al.\ 2006, ApJ, 650, 592

\bibitem[Lacki et al.(2010)]{lacki10} Lacki, B.~C., Thompson, T.~A., \& Quataert, E.\ 2010, ApJ, 717, 1 

\bibitem[Lanz et al.(2013)]{lanz13} Lanz, L., Zezas, A., Brassington, N., et al.\ 2013, ApJ, 768, 90 

\bibitem[Lau et al.(2015)]{lau2015} Lau, R.~M., Herter, T.~L., Morris, M.~R., Li, Z., \& Adams, J.~D.\ 2015, Science, 348, 413 

\bibitem[Lisenfeld \&V\"{o}lk(2010)]{lisenfeld10} Lisenfeld, U., V\"{o}lk, H.~J.\ 2010, A\&A, 524, AA27

\bibitem[Lisenfeld \&V\"{o}lk(2000)]{lisenfeld00} Lisenfeld, U., V\"{o}lk, H.~J.\ 2000, A\&A, 354, 423 

\bibitem[Lotz et al.(2008)]{lotz08} Lotz, J.~M., Jonsson, P., Cox, T.~J., \& Primack, J.~R.\ 2008, MNRAS, 391, 1137 

\bibitem[Magnelli et al.(2010)]{magnelli10} Magnelli, B., Lutz, D., Berta, S., et al.\ 2010, A\&A, 518, LL28 

\bibitem[Mazzarella et al.(2012)]{mazz12} Mazzarella, J.~M., Iwasawa, K., Vavilkin, T., et al.\ 2012, AJ, 144, 125

\bibitem[Mentuch Cooper et al.(2012)]{mc12} Mentuch Cooper, E., Wilson, C.~D., Foyle, K., et al.\ 2012, ApJ, 755, 165

\bibitem[Murphy et al.(2009)]{ematal09} Murphy, E.~J., Kenney, J.~D.~P., Helou, G., Chung, A., \& Howell, J.~H.\ 2009, ApJ, 694, 1435 

\bibitem[Murphy(2013)]{murphy13} Murphy, E.~J.\ 2013, ApJ, 777, 58

\bibitem[Petrosian et al.(1980)]{tigran80} Petrosian, A.~R., Saakian, K.~A., \& Khachikian, E.~E.\ 1980, Astrofizika, 16, 621 

\bibitem[Pettini et al.(1997)]{pettini1997} Pettini, M., King, D.~L., Smith, L.~J., \& Hunstead, R.~W.\ 1997, ApJ, 478, 536 

\bibitem[Prodanovi{\'c} et al.(2013)]{pub13} Prodanovi{\'c}, T., Bogdanovi{\'c}, T., \& Uro{\v s}evi{\'c}, D.\ 2013, PRD, 87, 103014 

\bibitem[Randriamampandry et al.(2015)]{ry15} Randriamampandry, S.~M., Crawford, S.~M., Cress, C.~M., et al.\ 2015, MNRAS, 447, 168 

\bibitem[Sajina et al.(2008)]{sajina08} Sajina, A., Yan, L., Lutz, D., et al.\ 2008, ApJ, 683, 659 

\bibitem[Sanders \& Mirabel(1996)]{sm96} Sanders, D.~B., \& Mirabel, I.~F.\ 1996, ARA\&A, 34, 749 

\bibitem[Sanders et al.(2003)]{sanders03} Sanders, D.~B., Mazzarella, J.~M., Kim, D.-C., Surace, J.~A., 
\& Soifer, B.~T.\ 2003, AJ, 126, 1607 

\bibitem[Sargent et al.(2010)]{sargent10} Sargent, M.~T., Schinnerer, E., Murphy, E., et al.\ 2010, ApJL, 714, L190 

\bibitem[Schinnerer et al.(2013)]{eva13} Schinnerer, E., Meidt, S.~E., Pety, J., et al.\ 2013, ApJ, 779, 42 

\bibitem[Schleicher \& Beck(2013)]{sb13} Schleicher, D.~R.~G., \& Beck, R.\ 2013, A\&A, 556, AA142 

\bibitem[Smolcic et al.(2014)]{smolcic14} Smolcic, V., Karim, A., Miettinen, O., et al.\ 2014, arXiv:1412.3799 

\bibitem[Spitzer (1958)]{spitzer58} Spitzer, L. J., 1958, ApJ, 127, 17

\bibitem[Strong et al.(2007)]{strong2007} Strong, A.~W., Moskalenko, I.~V., \& Ptuskin, V.~S.\ 2007, Annual Review of Nuclear and Particle Science, 57, 285 

\bibitem[Tammann et al.(1994)]{tammann94} Tammann, G.~A., Loeffler, W., \& Schroeder, A.\ 1994, ApJS, 92, 487 

\bibitem[Tielens(1998)]{tielens98} Tielens, A.~G.~G.~M.\ 1998, ApJ, 499, 267 

\bibitem[Thompson et al.(2006)]{thompson06} Thompson, T.~A., Quataert, E., Waxman, E., Murray, N., \& Martin, C.~L.\ 2006, ApJ, 645, 186 

\bibitem[van der Kruit(1971)]{vanderkruit71} van der Kruit, P.~C.\ 1971, A\&A, 15, 110 

\bibitem[Vardoulaki et al.(2015)]{var15} Vardoulaki, E., Charmandaris, V., Murphy, E.~J., et al.\ 2015, A\&A, 574, AA4

\bibitem[Villar-Mart{\'{\i}}n et al.(2001)]{villar-martin} Villar-Mart{\'{\i}}n, M., De Young, D., Alonso-Herrero, A., Allen, M., 
\& Binette, L.\ 2001, MNRAS, 328, 848 

\bibitem[Voelk(1989)]{voelk89} Voelk, H.~J.\ 1989, A\&A, 218, 67 

\bibitem[Yun \& Carilli(2002)]{yuncarilli02} Yun, M.~S., \& Carilli, C.~L.\ 2002, ApJ, 568, 88 

\bibitem[Yun et al.(2001)]{yun01} Yun, M.~S., Reddy, N.~A., \& Condon, J.~J.\ 2001, ApJ, 554, 803 

\end{thebibliography}
\end{document}